\documentstyle[11pt,newpasp,twoside,epsf]{article}
\def\edcomment#1{\iffalse\marginpar{\raggedright\sl#1\/}\else\relax\fi}
\reversemarginpar

\def\ltsima{$\; \buildrel < \over \sim \;$}
\def\lsim{\lower.5ex\hbox{\ltsima}}
\def\gtsima{$\; \buildrel > \over \sim \;$}
\def\gsim{\lower.5ex\hbox{\gtsima}}

\def\ltsima{$\; \buildrel < \over \sim \;$}
\def\gtsima{$\; \buildrel > \over \sim \;$}
\def\lsim{\lower.5ex\hbox{\ltsima}}
\def\gsim{\lower.5ex\hbox{\gtsima}}
\def\msun{${\rm M_\odot}~$}
\def\etal{{\it et al.}~}

\begin{document}

\title{Where Have All the Submillisecond Pulsars Gone?}

\author{L. Burderi, F. D'Antona, M.T. Menna, L. Stella}
\affil{Osservatorio Astronomico di Roma, Via Frascati 33, 
00040 Monteporzio Catone (Roma), Italy}
\author{A. Possenti, N. d'Amico, M. Burgay}
\affil{Osservatorio Astronomico di Bologna and Universit\`a 
di Bologna, Dipartimento di Astronomia, via Ranzani 1, 
40127 Bologna, Italy}
\author{T. Di Salvo}
\affil{Astronomical Institute ``Anton Pannekoek," University of 
Amsterdam,
Kruislaan 403, NL 1098 SJ Amsterdam, the Netherlands}
\author{R. Iaria, N.R. Robba}
\affil{Dipartimento di Scienze Fisiche ed Astronomiche, 
Universit\`a di Palermo, Via Archirafi 36, 90123 Palermo, Italy}
\author{S. Campana}
\affil{Osservatorio Astronomico di Milano}

\begin{abstract}
The existence of pulsars with spin period below one millisecond is
expected, though they have not been detected up to now. Their
formation depends on the quantity of matter accreted from the
companion which, in turn, is limited by the mechanism of mass ejection
from the binary. Mass ejection must be efficient, at least in some
cases, in order to produce the observed population of moderately fast
spinning millisecond pulsars. First we demonstrate, in the framework
of the widely accepted recycling scenario, using a population
synthesis approach, that a significant number of pulsars with spin
period below one millisecond is expected. Then we propose that
significant variations in the mass--transfer rate may cause, in systems
with orbital periods $\gsim 1$ h, the switch--on of a radio pulsar
whose radiation pressure is capable of ejecting out of the system most
of the matter transferred by the companion and prevent any further
accretion. We show how this mechanism could dramatically alter the
binary evolution since the mechanism that drives mass overflow from
the inner Lagrangian point is still active while the accretion is
inhibited. Moreover we demonstrate that the persistency of this
``radio ejection'' phase depends on the binary orbital period,
demonstrating that close systems (orbital periods $P_{\rm orb} \lsim
1$ h) are the only possible hosts for ultra fast spinning neutron
stars. This could explain why submillisecond pulsars have not been
detected so far, as current radio surveys are hampered by
computational limitations with respect to the detection of very short
spin period pulsars in short orbital period binaries.

\end{abstract}

\section{Introduction} 
 
PSR1937+21 is, at present, the neutron star (NS) having the shortest
rotational period $P_{\rm min}=1.558$ ms ever detected. Despite its
apparent smallness, $P_{\rm min}$ is not a critical period for NS
rotation: as shown by Cook, Shapiro \& Teukolsky (1994), the period
$P_{min}$ is longer than the limiting period, $P_{\rm lim}$ below
which the star becomes unstable to mass shedding at its equator.
Indeed $P_{\rm lim} \sim 0.5$ ms for most equations of state of the
ultra dense the nuclear matter, EoS). However, the re--acceleration of
a NS up to ultra short periods depends remarkably on the amount of
mass (and hence of angular momentum) accreted from a binary companion.
Back of the envelope calculation shows that $M_{\rm 1ms} \sim 0.2$
\msun must be accreted to attain $P \sim 1$ ms. Conservation of
angular momentum requires $M_{\rm 1ms}~l= I~ 2 \pi / ({\rm 1 ms})$ where
$I$ is the NS moment of inertia, $l=(GMR_{\rm acc})^{1/2}$ is the
specific angular momentum of the accreting matter, and $R_{\rm acc}$
is the accretion radius. Adopting $I=10^{45}$ g cm$^2$ and $R_{\rm
acc} \sim R_{\rm NS} = 10^6$ cm, where $R_{\rm NS}$ is the NS radius,
one gets the above result. More accurate calculation that includes
general relativistic effects and realistic EoS, have almost doubled
this estimate, showing that, typically $M_{\rm 1 ms} \sim 0.35$
\msun must be accreted to spin up a NS to  1 ms (Burderi \etal 1999).

Most donor stars in systems hosting recycled pulsars have certainly
lost, during their interacting binary evolution, a mass larger than
$M_{\rm 1 ms}$. They now appear as white dwarfs of mass $\sim 0.15-0.30$
\msun (Taam, King \& Ritter 2000), whose progenitors are likely to
have been stars of $\sim 1.0-2.0$ \msun (Webbink, Rappaport \&
Savonije 1983; Burderi, King \& Wynn 1996, Tauris \& Savonije
1999). Hence, a crucial parameter for attaining ultra fast rotation
is the fraction of the mass lost by the companion ($M_{\rm lost}=0.7-1.85$
\msun) which effectively accretes onto the NS which, in turns,
depends on the mechanisms driving the mass ejection from the
system.
 
In this paper we discuss the results of population synthesis
calculations for the formation of millisecond pulsars (MSPs) in the
framework of the standard recycling scenario (\S 2). We show that,
independent of the adopted EoS, the existence of a population of MSPs
spinning below 1 ms (submillisecond pulsar, SMSP) is predicted,
even though these systems have not been detected up to date.

In \S 3 we critically discuss the widely accepted recycling scenario for 
the formation of a MSP outlining the main difficulties of this model
by a comparison of the present observations with the results obtained 
in \S 2. 

To overcome these difficulties we propose, in \S 4, that, in some
cases, most of the transferred mass is ejected by the radiation
pressure generated by the rapidly spinning NS in a radio
pulsar regime. Moreover we investigate how the duration of this
process, and hence the amount of mass ejected, depends on the orbital
period of the binary system. The mechanism proposed (some aspects of
which were originally suggested by Ruderman, Shaham, \& Tavani 1989)
naturally provides an explanation for the observed values of masses
and spin periods of MSPs (Thorsett \& Chakrabarty 1999; Tauris \&
Savonije 1999). Our analysis indicates that only close
systems ($P_{\rm orb} \lsim 1$ h) could harbour SMSPs. In \S 5 we
discuss how this results in an observational bias that has prevented
the detection of SMSPs up to now.

\section{On the existence of submillisecond pulsars: a population synthesis 
approach}

Although $P_{\rm lim}$ depends only on the EoS adopted for the NSs,
the effective possibility to accelerate a NS to periods below 1 ms 
is determined by the modalities of the mass transfer from the binary 
companion and the evolution (decay) of the NS magnetic moment $\mu$ 
(Burderi \etal 1999).  
In \S 3 we discuss in some detail the possible scenarios for the 
mass transfer in these systems. 
No unique theory exists for the origin and evolution 
of the NS magnetic moment. 
The most credited theories suggest either that the field is a remnant 
from the progenitor star or that it has been generated soon after the 
formation of the NS.
In the first case, in which the electric currents generating the magnetic 
field are buried in the NS core, the evolution of $\mu$ is driven by 
secular variations in the rotational period of the NS during its evolution 
(e.g. Srinivasan \etal 1990; Miri \& Bhattacharya 1994; Ruderman 1998). 
In the other case, the electric currents are generated and confined in 
the thin crust of the NS and the magnetic field decay is driven by the 
ohmic diffusion (Sang \& Chanmugam 1987). In this latter case magnetic 
evolution is not univocal: as crustal matter is assimilated into the core 
during accretion, the magnetic field at the crust--core boundary can either
be expelled (boundary condition I, BC I: Urpin, Geppert \& Konenkov
1998) or advected into the superconductive core where it no longer
decays (BC II: Konar \& Bhattacharya 1999). Thus, at the endpoint of
evolution, the NS can either become almost non magnetic, or preserve a 
significant relic field. In this class of models, the magnetic field decay is
connected with the mass accretion history which determines both the amount
of accreted mass (which drives the advection process) and the thermal 
evolution of the crust (which modulates the ohmic dissipation process by 
influencing the electron and ion resistivity in the crust).

To explore all the possibilities mentioned above within the
recycling scenario for the formation of MSPs (Alpar \etal 1982; 
Bhattacharya 1995) Possenti \etal (1998, 1999) carried on
a statistical analysis of the NS spin period distributions in the 
millisecond and submillisecond range, using a population synthesis model
\footnote{The population synthesis code calculate the mass transfer
during the Roche lobe overflow considering a range of mass--transfer rates
which is close to the one observed in low mass X-ray binaries.
Eventually the population synthesis code follows the radio pulsar phase.  
The model incorporates the detailed physics of the evolution of a crustal 
magnetic field (using the two boundary conditions described in the text
to mimic expulsion or assimilation of the field by the NS core) and includes 
the relativistic corrections necessary to describe the spin--up process 
(Burderi \etal 1999). A Monte Carlo code (with 3,000 particles) is used to 
compute the sample population.}.

There is observational evidence (see Tanaka and Shibazaki 1996 for a
review) that NSs in low mass X-ray binaries (LMXBs) may undergo phases
of transient accretion (perhaps due to thermal viscous instabilities
in an irradiated disc e.g. van Paradijs 1996; King, Kolb, Burderi
1996).  This in turn can start one or more phases of {\it propeller}
which have been invoked to effectively spin down the NS and
eject most of the mass from the system, thus mitigating the effects of
the previous phases of spin--up. We will critically discuss the
propeller mechanism in \S 3 jeopardizing the whole propeller scenario.
In view of our discussion in \S 3, we investigate here the effects of
a propeller phase on the population of LMXBs by comparing two
possibilities: a persistent accretion for the whole duration of Roche
Lobe Overflow (RLO) phase, or a persistent accretion for half the
previous time followed by a quick stop of the accretion modeled as a
mass--transfer rate that varies with time $\rm \dot M \propto t^{\Gamma}$
with a power law index $\Gamma = 8$ representing a sudden drop
of the $\rm \dot M$. We assume that {\it all the transferred mass is ejected
from the system in this phase}.

\begin{table}
\scriptsize 
\centering
\vskip +0.2truecm
\begin{tabular}{llcr}
\tableline
\tableline
  &  &         &         \\
  & Statistical  & Values  &  Units \\         
  & Distribution &         &         \\  
  &  &         &         \\
\tableline
  &  &         &         \\
NS period at ${\rm t_0}$  &  Flat   & from 1  to  100  &   s \\
$\mu$ at ${\rm t_0}$  & Gaussian in Log & 
$Log<\mu_0>=28.50;~\sigma=0.32$ &  $G~cm^{3}$ \\
${\dot m}$  & Gaussian in Log & 
$Log<{\rm \dot m}>=-1.00;~\sigma=0.5$ & $\rm \dot M_E$ \\
Minimum accreted mass & One-value & 0.01 & ${\rm {M_{\odot}}}$ \\
Maximum accreted mass & One-value & 0.5  & ${\rm {M_{\odot}}}$ \\
duration of RLO phase & Flat in Log & 
from $10^6$ to $5 \times 10^{8} $ & year \\
duration of MSP phase &  Flat in Log & 
from $10^8$ to $3 \times 10^{9} $ & year \\
  &  &         &         \\
\tableline
\tableline
\end{tabular}
\caption{\footnotesize Population synthesis parameters. 
${\rm t_0}$ is initial time of the RLO phase and $\rm \dot M_E$ is 
the Eddignton accretion rate.}
\label{tab:input_values_pop}  
\end{table}

\begin{figure}
\vspace*{-0.1in}
\plottwo{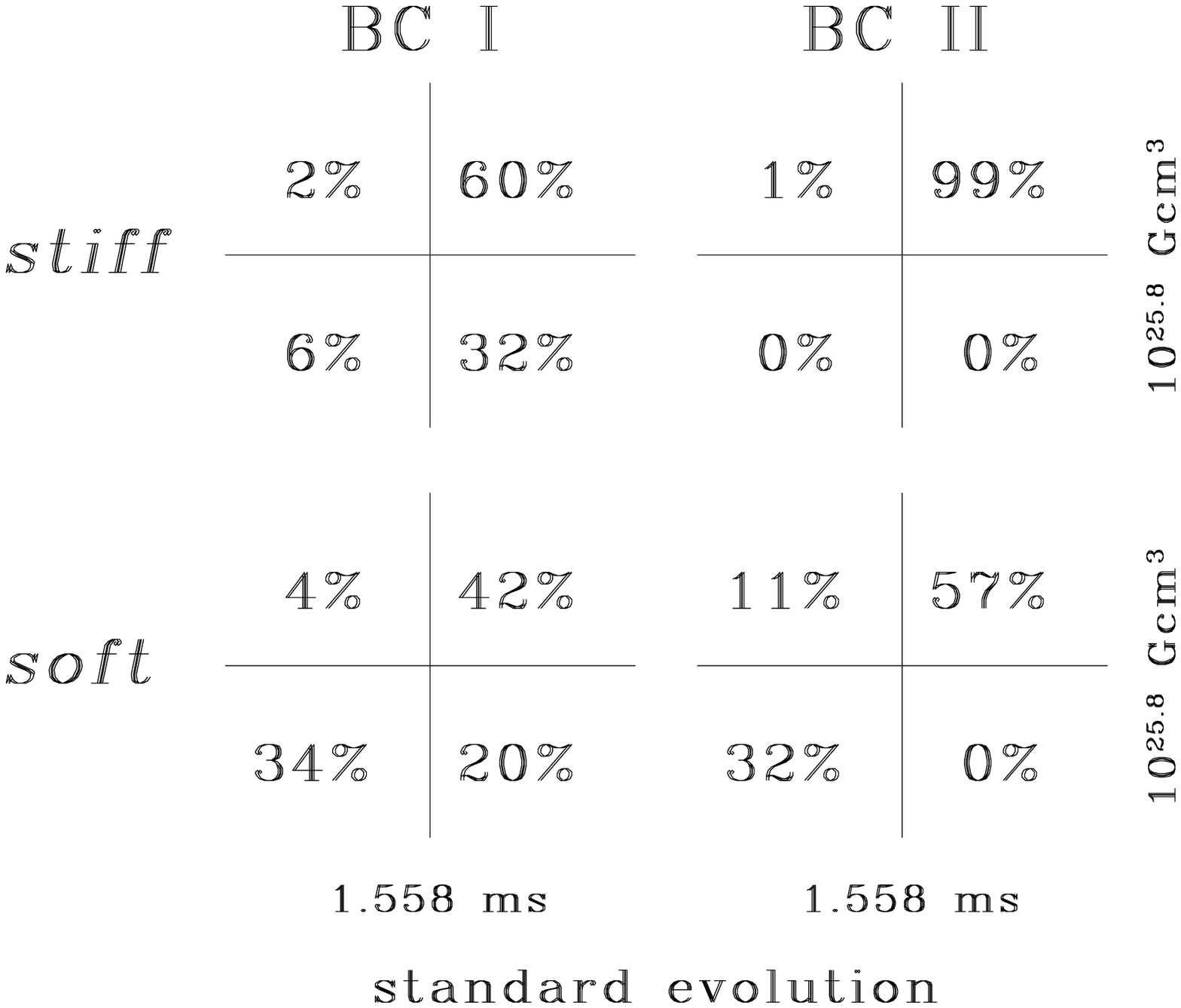}{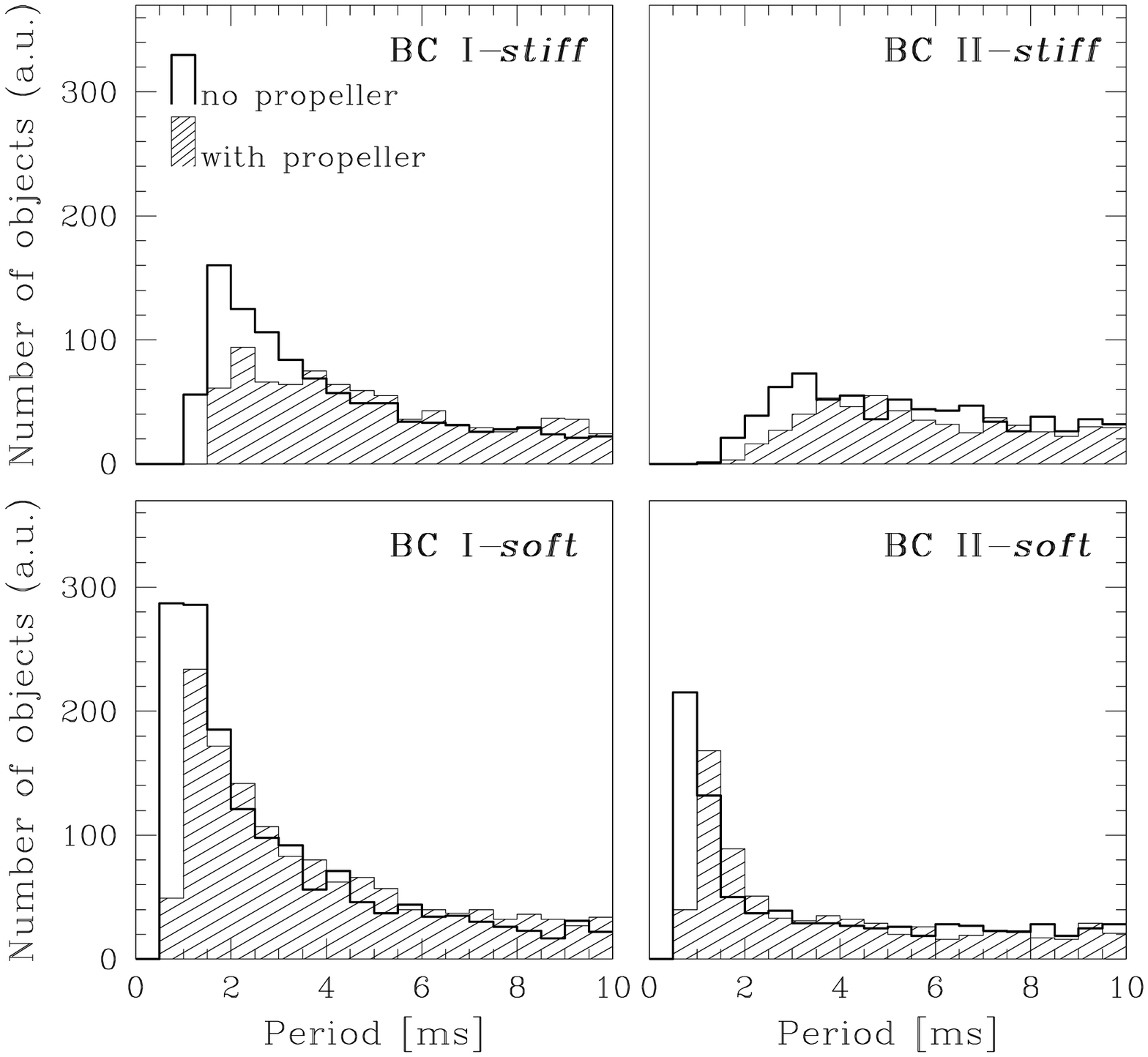}
\caption{\label{fig:crosses} 
Left: percentage distributions of the synthesized MSPs. 
The four panels indicate, from upper right and counterclockwise: 
a) objects having $P>P_{\rm min}$ and $\mu>\mu_{\rm min}$; 
b) $P<P_{\rm min}$ and $\mu>\mu_{\rm min}$; 
c) $P<P_{\rm min}$ and $\mu<\mu_{\rm min}$; 
d) $P>P_{\rm min}$ and $\mu<\mu_{\rm min}$; 
(the typical variance is about 1\%). 
Right: spin period distributions of the simulated NS ($\mu$ varies
over the whole range). The solid lines indicates the distributions
in absence of propeller, the dashed areas represent the distributions 
with a strong propeller effect. The total number of
objects is in arbitrary units (a.u.).}
\end{figure}

The left panels in figure \ref{fig:crosses} give the percentage distribution
of the recycled NSs synthesized using the set of parameters reported
in table \ref{tab:input_values_pop}. 
Adopting as references the values of
$P_{\rm min}$ and $\mu_{\rm min}$ (the shortest rotational period observed in
PSR~1937+21 and the weakest magnetic moment observed in
PSRJ2317+1439), the NS are divided into four groups: \\
\begin{itemize}
\item[a)] $P\geq P_{\rm min}$ and $\mu\geq\mu_{\rm min}$ these resemble the 
MSPs that have been observed so far; 
\item[b)] $P<P_{\rm min}$ and $\mu\geq\mu_{\rm min}$ these objects represent a
population of ultra fast radio pulsars (see Burderi \& D'Amico 1997); 
\item[c)] $P<P_{\rm min}$ and $\mu < \mu_{\rm min}$ ultra fast radio 
pulsars with a weak magnetic field. Indeed most of these objects will be 
above the Chen and Ruderman (1993) ``death--line'', and might have a bolometric
luminosity comparable to that of the known MSPs. For simplicity,
hereafter we refer to all the objects having $P<P_{\rm min}$ and
$\mu$ above the ``death--line'' as SMSPs; 
\item[d)] $P\geq P_{\rm min}$ and $\mu < \mu_{\rm min}$ these are 
probably radio quiet NSs because they are close or below the theoretical 
``death--line'' (actually this period range was already searched with 
negative results by present, good sensitivity radio surveys).
\end{itemize}
The four panels 
represent two EoSs (a very stiff one corresponding to
a minimum spin period $P_{\rm lim} \simeq 1.4$ ms and a mildly soft one,
with $P_{\rm lim} \simeq 0.7$ ms) and the two different boundary conditions
discussed above for the magnetic field at the
crust--core interface (BC I \& BC II).

It appears that objects with periods $P<P_{\rm min}$ are present in a
statistically significant number.  In effect, {\it a tail of}
SMSPs {\it is always present for any reasonable choice of the
parameters} in table \ref{tab:input_values_pop}. 
The right panels of figure \ref{fig:crosses} show the distributions of the 
number of objects {\it vs} the spin period for the NS of our synthetic sample.

We first discuss the results in absence of a propeller phase (solid
lines).  For the mildly soft EoS, the ``barrier'' at $P_{\rm lim}
\simeq 0.7$ ms is clearly visible as the mildly soft EoS produces
distributions that increase rather steeply towards periods smaller
than 2 ms, irrespective of the boundary condition adopted for $\mu$.
On the other hand, the boundary conditions on $\mu$ affect the
fraction of objects with short periods since the magnetospheric radius
(that determines the equilibrium spin period) depends on the magnetic
moment (see \S 3).  BC II produces a smaller number of objects with a
low field as the field initially decays but, when the currents are
advected towards the crust--core boundary, the decay is halted and the
field reaches a bottom value. Even the very stiff EoS permits periods
$P<P_{\rm min}$, but the ``barrier'' of mass shedding (at $P_{\rm
lim}\simeq 1.4$ ms) is so close to $P_{\rm min}=1.558$ ms that only
few NSs reach these extreme rotational rates. 

The effects of a strong propeller phase (lasting $\sim50\%$ of the RLO phase 
with an efficiency in ejecting matter from the system close to $100\%$ -- 
although this assumption will be strongly questioned in \S 3) are shown 
by the dashed areas.
We note that {\it significant mass losses from the system can threaten
the formation of NSs with} $P < P_{\rm min}$ only in the (unlikely) case  of 
the very stiff EoS.

Possenti \etal (1998, 1999) computed the distributions of the number of 
objects {\it vs} the final masses 
for the different spin periods attained at the end of the evolution. 
The distributions steepens towards high mass values (approaching 
$M\sim 1.7 \div 1.8~M_{\odot}$) for final spin periods below 
$\sim P_{\rm min}$. 
All this results are in line with a general conclusion, namely: 
{\it an accretion of about $0.35$ \msun is a general necessary condition to 
spin a NS to ultra short periods} (see Burderi \etal 1999 for a more technical 
discussion that takes into account general relativistic effects).

\section{Recycling at sub--Eddington rates} 

In line with a widely used convention let us define {\it primary} the NS
and {\it secondary} its companion.

It is well known that an upper limit to the accretion rate is given by
the Eddington limit, but for typical companion initial masses $M_{\rm ini}
\lsim 1.6$ \msun and initial $P_{\rm orb}\lsim 50$ days,
the donor transfers mass at sub--Eddington rates, making, in principle,
the whole mass lost by the secondary available for the recycling of the NS. 

\begin{table}
\scriptsize 
\centering
\vskip +0.2truecm
\begin{tabular}{lcccc}
\tableline
\tableline
  &  &         &    &   \\
& System n. & {${M_{NS}}({\rm M_\odot})$} & 
{${M_{\rm donor}}({\rm M_\odot})$} & {${P_{\rm orb}}({\rm h})$} \\
   &  & & & \\
\tableline
   &  & & & \\
Initial & 1 & 1.40 & 1.20 & 10.46 \\ 
Final   &   & 2.58 & 0.02 & 1.42 \\
   &  & & & \\
Initial & 2 & 1.40 & 1.20 & 18.64 \\ 
Final   &   & 2.39 & 0.21 & 92.15 \\ 
   &  & & & \\ 
Initial & 3 & 1.40 & 1.20  & 30.55 \\ 
Final   &   & 2.35 & 0.25  & 338.1   \\ 
   &  & & & \\ 
Initial & 4 & 1.40 & 1.60 & 34.00 \\ 
Final   &   & 2.67 & 0.33 & 346.0 \\ 
   &  & & & \\
\tableline  
\tableline
\end{tabular}
\caption{ \footnotesize Mass and period evolution in some LMXBs}
\label{tab:dantona}  
\end{table} 
 
\begin{figure}
\vspace*{3.5cm}
\plotfiddle{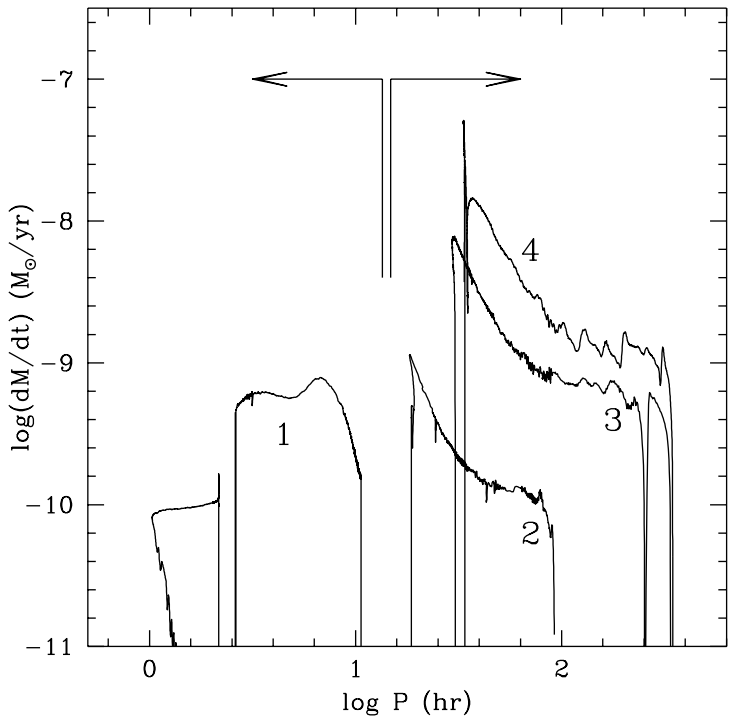}{5. cm}{0}{50}{50}{-150}{-85}
\caption{\label{fig:dantona} $\dot M$ {\it vs} $P_{\rm orb}$
for the systems of table \ref{tab:dantona}. 
The temporary stop of $\dot M$ in track 1 is due to the switch--off of the
magnetic braking once the star becomes fully convective.  The sharp
drop of $\dot M$ in track 3 is due to a chemical readjustment of the
star structure.  The arrows show the $P_{\rm orb}$ that separates
cases A and B of mass transfer for a $1.2$ \msun secondary.}
\end{figure}

We explored this scenario ({\it i.e.} conservative mass transfer) computing 
the system evolution (with initial parameters as in table \ref{tab:dantona}) 
with the ATON1.2 code (D'Antona, Mazzitelli \& Ritter 1989). The mass loss 
rate by the companion {\it vs.} the orbital period is shown in figure 
\ref{fig:dantona}. The mass loss rate was computed, following the formulation 
by Ritter (1988), as an exponential function of the distance between the 
stellar radius and the secondary Roche lobe in units of the pressure scale 
height. 
This method also allows to compute the first phases of mass transfer,
during which the rate reaches values which can be much larger than the
stationary values due to the thermal response of the star to mass
loss.

Besides the cases of systems with long $P_{\rm orb}$ in which the donor fills
the Roche lobe while it is evolving towards the red giant branch (case
B of mass transfer, systems n.2,3,4), we consider the case of a system with 
short $P_{\rm orb}$ and a secondary of 1.2 \msun that fills the
Roche lobe during the core--hydrogen burning phase (case A of mass transfer, 
system n.1). In systems with long $P_{\rm orb}$, mass transfer is driven by 
the thermal and nuclear evolution of the secondary; in systems with short 
$P_{\rm orb}$ systems mass transfer is driven by magnetic braking acting 
until $P \simeq 2.5$ h, when the star becomes fully convective. 
The mass transfer then stops and resumes at a shorter period driven by 
losses of angular momentum by gravitational radiation, like in  
cataclysmic variables. 
In this case the minimum period attainable is $\sim 1.03$ h. 
On the other hand in the systems with long $P_{\rm orb}$ the mass transfer 
terminates with periods in the 90 -- 400 h range.

In the systems with long $P_{\rm orb}$ the final masses of the donor stars 
are in the $0.21-0.33$ \msun range, in agreement with the masses measured 
for the companions of MSP, as inferred from accurate timing campaigns 
(see Burderi, King, \& Wynn 1996). However, the predicted final masses of the NSs are 
in the $2.35-2.67$ \msun range, very close to (or even bigger than) the 
maximum mass allowed for a NS in most of the proposed EoSs (Cook, Shapiro \& 
Teukolsky 1994). This suggests either that mass transfer cannot be 
conservative -- and most of the donor mass must be expelled from the 
system -- or that the final NS must be very massive implying that 
accretion--induced collapse to a black hole is a likely outcome for LMXBs.

The propeller effect (Illarionov \& Sunyaev 1975) has often been invoked
as the process capable of expelling most of the mass transferred by the
companion star. 
The accretion disc is truncated at
the magnetospheric radius $R_{\rm m}$, at which the magnetic field
pressure equals the pressure of the matter in the disc
(see {\it e.g.} Hayakawa 1985 for a review). For a
Shakura--Sunyaev accretion disc an expression of $R_{\rm m}$ can be
derived ({\it e.g.}  Burderi \etal 1998)
\begin{eqnarray} 
R_{\rm m} & = & 4.70 \times 10^{5} \alpha^{4/15} \epsilon^{34/135} 
n_{0.615}^{8/27} L_{37}^{-34/135} m^{-1/135} \nonumber \\  
            &   & R_6^{-34/135} f^{-136/135} \mu_{26}^{16/27}   
\;\; {\rm cm}  
\label{eq:rm} 
\end{eqnarray}  
where $\alpha$ is the Shakura--Sunyaev viscosity parameter, 
$\epsilon \sim 1$ is the ratio between the observed luminosity and the total  
gravitational potential energy released per second by the accreting matter, 
$n_{0.615} = n/0.615$ $\sim 1$ for a gas with solar abundances (where $n$ is  
the mean particle mass in units of the proton mass $m_{\rm p}$), 
$L_{37}$ is the accretion luminosity in units of $10^{37}$ erg/s
that measures $\dot M$ in the hypothesis that {\it all the transferred mass
accretes onto the NS} radiating a specific energy 
$\epsilon \times GM/R_{\rm NS}$, 
$m$ is the NS mass $M$ in solar masses, 
$R_6$ is the NS radius in units of $10^6$ cm, 
$f = [1 - (R_6/R_{\rm m})^{1/2}]^{1/4} \sim 1$, 
$\mu_{26}$ is the magnetic moment of the NS in units of 
$10^{26}$ G cm$^3$ ($\mu = B_{\rm s} R^3$ where $R$ and $B_{\rm s}$ 
are the NS radius and surface magnetic field along the 
magnetic axis respectively). At radii $r < R_{\rm m}$ the accreting matter is 
forced by the magnetic field to corotate with the NS and  
accretes along the field lines.  As $R_{\rm m} \propto L_{37}^{-34/135}$,
a decrease in the mass--transfer rate results in an 
expansion of the magnetosphere. Accretion onto a spinning magnetized 
NS is centrifugally inhibited once $R_{\rm m}$ lies outside 
the corotation radius $R_{\rm CO}$, the radius at which the Keplerian 
angular frequency of the orbiting matter is equal to the NS 
spin: 
\begin{equation} 
R_{\rm CO} = 1.50 \times 10^{6} m^{1/3} P_{-3}^{2/3} \;\; cm  
\label{eq:rco} 
\end{equation} 
where $P_{-3}$ is the spin period in milliseconds.  In this case a
significant fraction of the accreting matter might be centrifugally
ejected from the system: this is the so called propeller phase.

The virial theorem sets stringent limits on the fraction of matter that can be
ejected in this phase, in fact, it states that, at any radius
in the disc, the
virialized matter has already liberated ({\it via} electromagnetic radiation)
half of its available energy, corresponding to the variation of its
potential energy.  Considering that $R_{\rm m} \sim R_{\rm NS} $ for
MSPs, the matter at the magnetosphere has irradiated $\sim 50\%$ of
the whole specific energy obtainable from accretion $\eta_{\rm acc} =
GM / R_{\rm NS}$.  To eject the same matter (that is close to the
NS surface) $\sim 1/2 ~\eta_{\rm acc} $ must be given back to it.  As
the only source of energy is the accretion process itself, the maximum
ejection efficiency is $\leq 50 \%$.  For instance, with a fine tuned
alternation of accretion and propeller phases of the right duration,
the kinetic energy stored by the NS in the form of rotational energy
during an accretion phase, can be used in a subsequent propeller phase
to eject almost the same amount of matter previously accreted.  The
major difficulty with this scenario is that, during the accretion
phase, once the system has reached the spin--equilibrium, no further
spin up takes place, and thus the storage of accretion energy in a
form that allows its subsequent re--usage for ejection is
impossible.  This is the reason why the duration of the accretion and
propeller phases has to be {\it ad hoc} chosen in order to reach
ejection efficiencies close to 50 \%.  It follows that, {\it in any
case, the accreted mass is no less than half of the transferred mass,
i.e.}  $\ge 0.4-0.8$ \msun.  This has two main consequences:
\begin{itemize}
\item[1)] even taking a propeller phase
into account, NSs in MSPs will either be very massive or even collapse into
black holes; 
\item[2)] as the amount of mass accreted is considerably
bigger than the minimum required to spin up the NSs to ultra short
periods (Burderi \etal 1999), one has to invoke an {\it ad hoc} final,
long lasting propeller phase with an highly effective spin down
to form the observed population of moderately fast spinning MSPs. 
\end{itemize}

The only way to overcome these difficulties is to obtain ejection efficiencies
close to unity. This is indeed possible if matter is ejected far from the NS 
surface through a large expansion of the magnetospheric radius 
($R_{\rm m} \rightarrow r >> R_{\rm NS}$), at a distance where the orbiting 
matter has an almost negligible binding energy $ GM / r << \eta_{\rm acc}$. 
As the NS is spinning very fast, the switch on of a radio pulsar is unavoidable
once $R_{\rm m} \rightarrow r$ (see \S 4).
In the next section we demonstrate that, under particular circumstances,
the pressure exerted by the radiation field of the radio pulsar overcomes
the pressure of the accretion disc, thus determining the ejection of the
matter from the system. We note that, once the disc has been swept away,
the radiation pressure stops the infall of matter far away from the NS surface,
{\it i.e.} in a region where $\eta_{\rm acc} \sim 0$, thus overcoming the 
50 \% efficiency limit imposed by the virial theorem.
This means that the pressure of the radiation emitted by pulsar is a viable 
mechanism to 
sweep most of the matter away from the system. We discuss this possibility in 
the next section.

\section{The effects of the pulsar energy outflow} 
 
The interaction of the accreting matter with the (assumed dipolar)
magnetic field of the NS can be described in terms of an outward
pressure (we use the expressions {\it outward} or {\it inward
pressures} to indicate the versus of the force with respect to the
radial direction) exerted by the rotating magnetic field on the
accretion flow (see {\it e.g.}  Hayakawa, 1985).

The light--cylinder radius, $R_{\rm LC} = c P / 2 \pi$ (where an
object corotating with the NS reaches the speed of light $c$),
separates an inner region, where the magnetic dipole pressure term is
relevant, from an outer region where the radiation pressure of the
rotating dipole is acting.

\begin{figure}
\plotfiddle{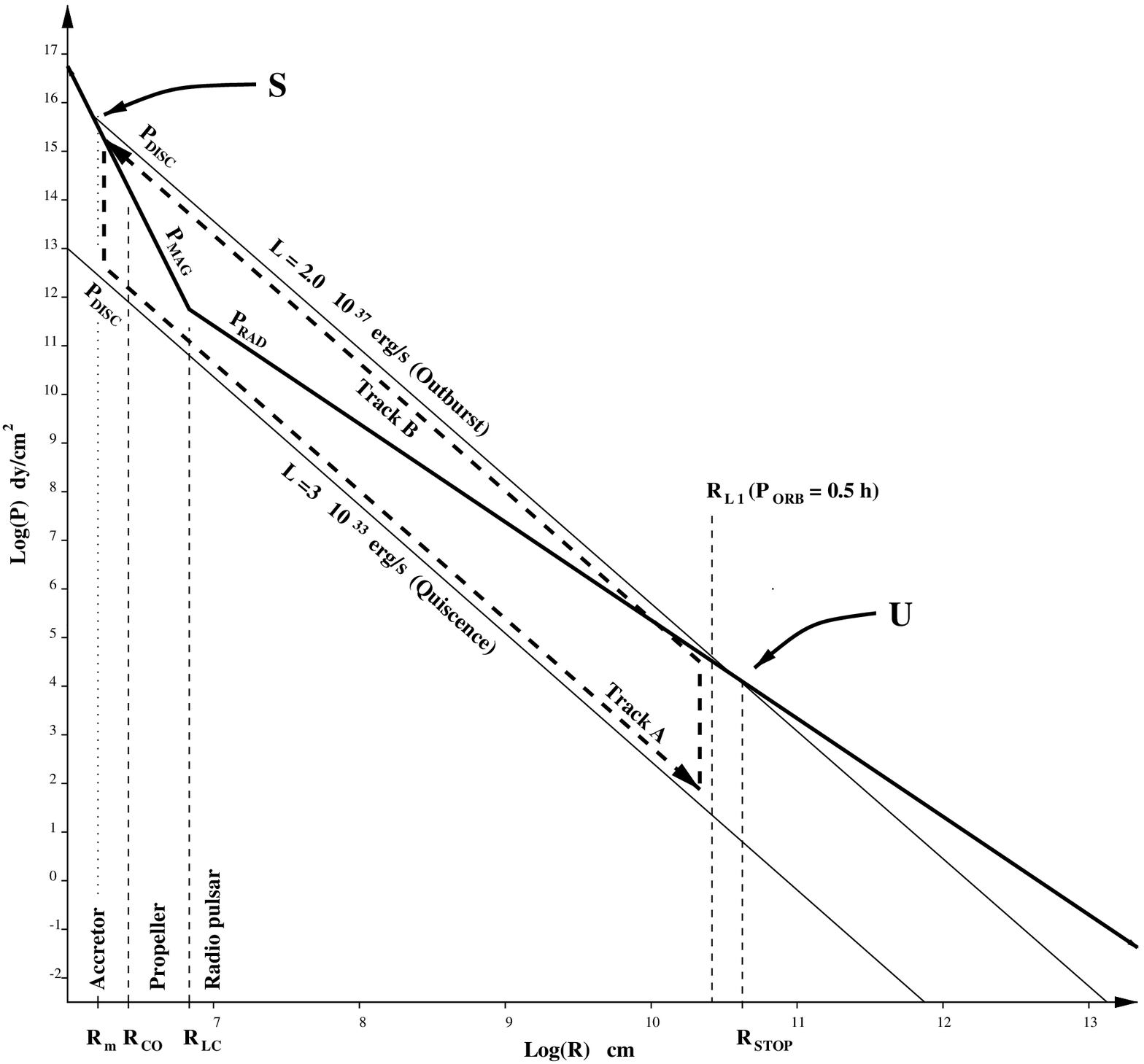}{10. cm}{0}{60}{60}{-165}{0}
\plotfiddle{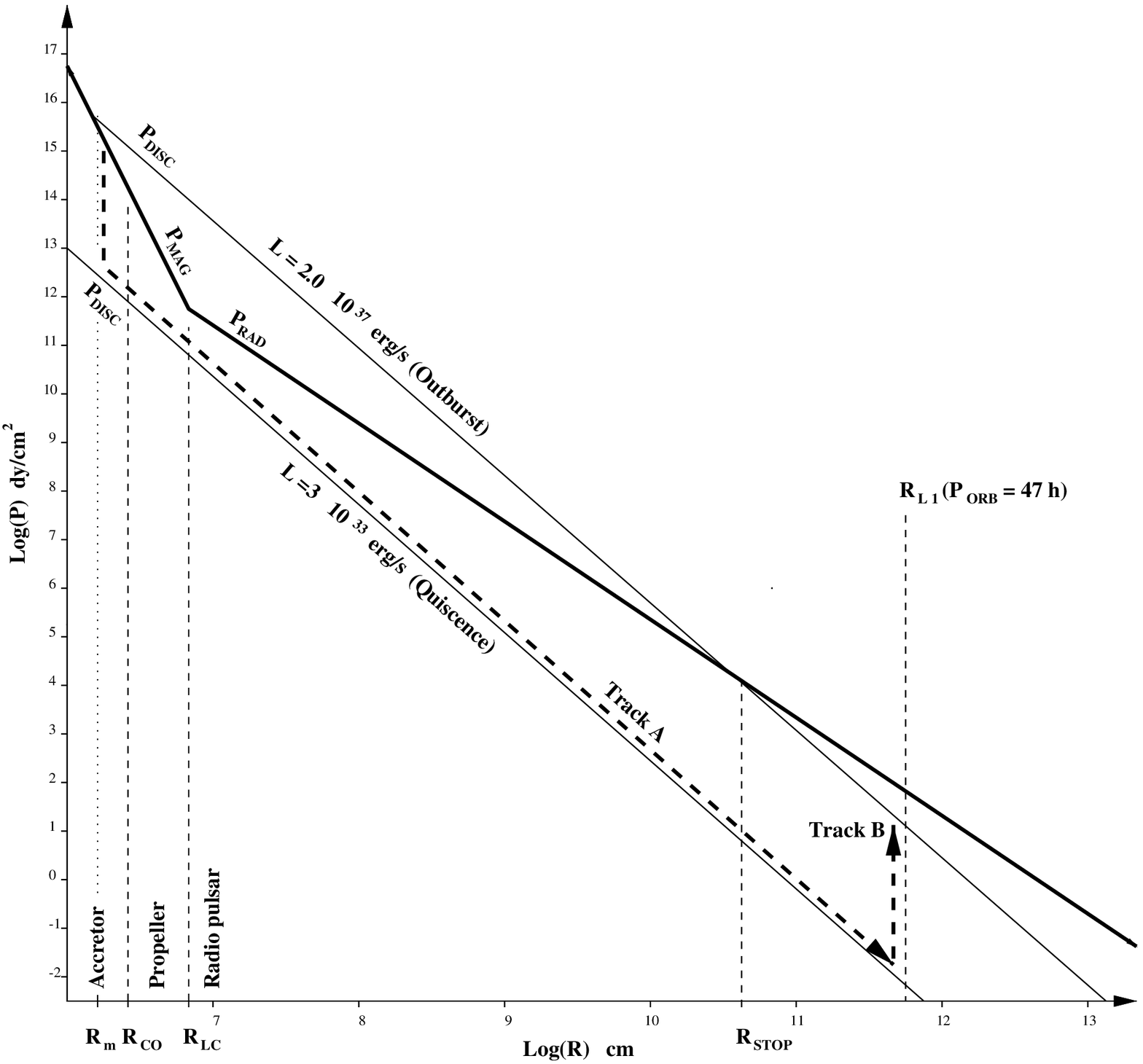}{10. cm}{0}{60}{60}{-165}{0}
\caption{\label{fig:pressure} The radial dependence of the pressures
relevant for the evolution of accreting NSs and recycled pulsars for:
a compact system ($P_{\rm orb} = 0.5$ h), upper panel, and a wide
systems ($P_{\rm orb} = 47$ h), lower panel.  The parameters adopted
are: $\mu_{26} = 7.7$, $P_{-3} = 1.5$, $\alpha = 1$, $\epsilon = 1$,
$n_{0.615} =1$, $R_6 = 1$, $m = 1.65$.}
\end{figure}

For $r < R_{\rm LC}$, the outward pressure, due to the magnetic field, is
\begin{equation}
P_{\rm MAG} = \frac{B^2}{8\pi} = 7.96 \times 10^{14} \mu_{26}^2 r_6^{-6} 
\;\; {\rm dy}/{\rm cm}^2
\label{eq:pmag}
\end{equation}
where $r_6$ is the distance from the NS center in units of $10^6$ cm.
For $r > R_{\rm LC}$ the outward pressure is given by the radiation
pressure which, assuming isotropic emission, is given by:
\begin{equation}
P_{\rm RAD} = 2.04 \times 10^{12} P_{-3}^{-4} 
\mu_{26}^2 r_6^{-2} \;\; {\rm dy}/{\rm cm}^2.
\label{eq:prad}
\end{equation}
In figure \ref{fig:pressure} the two outward pressures defined above
are shown as bold lines for typical values of the parameters (see
figure caption).  
 
The accretion flow, in turns, exerts a inward pressure on the field
that is the sum of the thermal gas pressure $P_{\rm gas} = \rho kT / n
m_{\rm p}$, and the light radiation pressure $P_{\rm light} = \sigma
T^4 / c$, where $\rho$ and $T$ are the flow density and temperature,
$k$ is the Boltzmann constant, and $\sigma$ is the Stefan--Boltzmann
constant. In the case of a Shakura--Sunyaev accretion disc orbiting a
compact object of $\sim 1$ \msun we have $P_{\rm light} \lsim P_{\rm
gas}$ for $\dot M < {\rm \dot M_{E}}$ and $r > R_{\rm NS}$ (see {\it
e.g.} Frank, King \& Raine 1992). Adopting for the inward disc presure
$P_{\rm DISC} \simeq P_{\rm gas}$ we have
\begin{eqnarray} 
P_{\rm DISC} & = & 1.02 \times 10^{16} \alpha^{-9/10} \epsilon^{-17/20} 
n_{0.615}^{-1} L_{37}^{17/20} m^{1/40} \nonumber \\   
            &   & R_6^{17/20} f^{17/5} r_6^{-21/8}  
\;\; {\rm dy}/{\rm cm}^2  
\label{eq:pgas} 
\end{eqnarray}   
where $f=[1-(R_6/r_6)^{1/2}]^{1/4} \leq 1$ and $r_6$ is the generic
radial distance in units of $10^6$ cm.  As before we measure $\dot M$
in units of $L_{37}$ in the hypothesis that {\it all the transferred
mass accretes onto the NS} radiating a specific energy $\epsilon
\times GM/R_{\rm NS}$.

The disc pressure line tangent to the intersection point of
(\ref{eq:pmag}) and (\ref{eq:prad}) defines a critical mass--transfer
rate $M_{\rm switch}$ ({\it i.e.} a critical luminosity $L_{\rm
switch}$ ) at which the radio pulsar switches--on. This depends on the
position of the light--cylinder radius, and hence on the spin period
of the NS: the faster is the rotation of the NS, the smaller is
$R_{\rm LC}$, and the larger is the critical mass--transfer rate at
which the pulsar switches--on.

In figure \ref{fig:pressure} the inward disc pressures (equation
\ref{eq:pgas}) for two different $\dot M$ (measured in units of
$L_{37}$ as discussed above) are shown as solid lines.  The disc
pressure corresponding to $L_{\rm switch}$ is shown as a dotted line.

For a given mass--transfer rate $\dot M_{\rm high}$ (corresponding to a
luminosity $L = 2 \times 10^{37}$ erg/s in figure \ref{fig:pressure}),
the intersections of the relative $P_{\rm DISC}$ line with each of the
inward pressure lines (points S and U in the upper panel of the same
figure) define equilibria points between the inward and outward
pressures. However, the equilibrium is {\it stable} at S ({\it i.e.}
at $r = R_{\rm m} $, as defined by equation \ref{eq:rm}), and {\it
unstable} at U ({\it i.e.} at $r = R_{\rm STOP}).$\footnote {An explicit 
expression for $R_{\rm STOP}$ can be derived equating
(\ref{eq:prad}) and (\ref{eq:pgas}).
\begin{equation} 
R_{\rm STOP} =  8.26 \times 10^{11} \alpha^{-36/25} \epsilon^{-34/25} 
n_{0.615}^{-8/5} L_{37}^{34/25} m^{1/25}  R_6^{34/25} f^{136/25} \mu_{26}^{-16/5} P_{-3}^{32/5}   
\;\; {\rm cm}.  
\label{eq:rstop} 
\end{equation}
}

In fact, as $P_{\rm MAG}$ is steeper than $P_{\rm DISC}$\ at S, if a
small fluctuation forces the inner rim of the disc inward (outward),
in a region where the magnetic pressure is greater (smaller) than the
disc pressure, this results in a net force that pushes the disc back
to its original location $R_{\rm m}$. On the other hand, as $P_{\rm
RAD}$ is flatter than $P_{\rm DISC}$\ at U, no stable equilibrium is
possible at $R_{\rm STOP}$. In fact, if a small fluctuation forces the
inner rim of the disc inward, $P_{\rm DISC}$ dominates over $P_{\rm
RAD}$ and the disc moves inward until the stable equilibrium point S
at $R_{\rm m}$ is reached. But, conversely, if a small fluctuation
forces the inner rim of the disc outward, $P_{\rm RAD}$ dominates over
$P_{\rm DISC}$ and the disc is swept away by the radiation
pressure. This means that, for $r>R_{\rm STOP}$\ no disc can exist for
any mass--transfer rate $\leq \dot M_{\rm high}$.

It is convenient to define {\it compact} systems those in which the
primary Roche lobe radius ($R_{\rm L 1}\propto P_{\rm orb}^{2/3}$)
lies inward $R_{\rm STOP}$ (figure \ref{fig:pressure}, upper panel)
and {\it wide} those in which the primary Roche lobe radius lies
outward $R_{\rm STOP}$ (figure \ref{fig:pressure}, lower panel), as
these systems behave very differently in response to significant and
abrupt variations of the mass--transfer rate. 

Let us consider, indeed,
the effects of such a sudden variation in the mass--transfer rate, as
is the case {\it e.g.} of a transient system alternating between
outbursts ($\dot M_{\rm high}$) and quiescence ($\dot M_{\rm low}$)
phases. The corresponding luminosities are $2 \times 10^{37}$ and $3
\times 10^{33}$ erg/sec, respectively. 

For simplicity, in the transition from outburst to quiescence
(quiescence to outburst), let us assume that the timescale on which
the drop (rise) in the disc pressure occurs is much shorter than the
timescale on which the magnetosphere expands (contracts), as this does
not affect the general conclusions.

The behaviour of a compact system (short orbital period, {\it e.g.}
$P_{\rm orb} = 0.5$ h) is cyclic, as shown in figure
\ref{fig:pressure}, upper panel.
\begin{itemize}
\item[Ic] {\it Accretor.}  
During the outburst, the magnetospheric radius (delimiting the inner
radius of the disc) is smaller than both the corotation and the
light--cylinder radius and the NS will normally accrete matter and
angular momentum, thus increasing its spin.

\item[IIc] {\it Radio ejector.}  The sudden drop in the mass--transfer
rate (quiescence) initiates a phase, that we termed ``radio
ejection'', in which the mechanism that drives {\it mass overflow}
from the inner Lagrangian point L 1 is still active, while the pulsar
radiation pressure at L 1 prevents mass accretion. In fact, as soon as
the new disc pressure (at the same outburst magnetospheric radius,
because of our assumption on the drop and rise timescales) drops below
$P_{\rm switch}$, the magnetosphere expands beyond the light cylinder
radius and the radio pulsar switches--on pushing the disc away from
the system until $R_{\rm L 1}$ is reached starting the radio ejection
phase (track A). As the overflowing matter cannot accrete, it is now
ejected as soon as it enters the Roche lobe of the primary, where its
binding energy is negligible, thus circumventing the limit imposed by
the virial theorem discussed in the previous section and allowing
ejection efficiencies close to $100\%$. The ejected matter presumably
leaves the system with a specific angular momentum $l$ between those
of the inner and outer Lagrangian points $(2 \pi /P_{\rm orb}) d_{\rm
L 1}^2 < l < (2 \pi /P_{\rm orb}) d_{\rm L 2}^2$, where $d_{\rm L 1}$
and $d_{\rm L 2}$ are the distances between the centre of mass of the
system and the inner and outer Lagrangian points respectively.

\item[Ic] {\it Accretor.} 
When a new outburst occurs, the mass--transfer rate rises back to its
original value ($L = 2 \times 10^{37}$ ergs/s), the disc follows track
B, and the accretion resumes. In the subsequent quiescence, the
system goes back to a radio ejection phase IIc, alternating between Ic
and IIc in response to the variations of the mass--transfer rate.
\end{itemize}

The response of a wide system (long orbital period, {\it e.g.}  47 h)
with the same transient behaviour is quite different, as shown in
figure \ref{fig:pressure}, lower panel.
\begin{itemize}
\item[Iw] {\it Accretor.}
For the same initial mass--transfer rates as in Ic, the NS accretes matter 
and increases its spin.
\item[IIw] {\it Radio ejector.}
The same drop in mass--transfer as in case IIc triggers the  
switching--on of the radio pulsar and the subsequent radio ejection of the
matter overflowing $R_{\rm L 1}$ (track A). 
\item[IIIw] {\it Radio ejector.}  Contrary to the compact system case,
when the mass--transfer rate recovers its previous value, the cyclic
behaviour is lost. Indeed, since for wide systems $R_{\rm L 1}$ (that
is the endpoint of track A for both wide and compact systems) is
located beyond $R_{\rm STOP}$, $P_{\rm DISC} < P_{\rm RAD}$ even for
$L = 2 \times 10^{37}$ ergs/s and the accretion can not resume.  The
growth of pressure indicated by track B is not sufficient to allow
disc formation.  This means that for a wide system once a drop of the
mass--transfer rate has started the radio ejection, a subsequent
restoration of the original mass--transfer rate is unable to quench the
ejection process (as already pointed out by Ruderman, Shaham, \&
Tavani 1989).
\end{itemize}
 
We note that the radio ejection mechanism proposed here can also work
if the variations of the mass--transfer rate are determined by the secular
evolution rather than by a transient behaviour.  Especially in this
case, once the situation described in cases Iw, IIw, \& IIIw occurs,
the subsequent orbital evolution is very sensitive to the exact value
of $l$ and will be discussed elsewhere (Burderi \etal, 2001, in
preparation) leading in some cases to the disruption of the companion.
This disruption mechanism (let us call it ``Roche lobe disruption'')
is in many respects similar to the effect of the ``evaporation'' of
the companion irradiated by the wind (composed of electromagnetic
radiation and relativistic particles) of a MSP (Tavani 1991). However
several differences exists that will be discussed elsewhere (Burderi
\etal, 2001, in preparation).

In conclusion, while the evolution without a radio ejection phase
implies that a large fraction of the transferred mass is accreted onto
the NS (because of the constraints imposed by the virial theorem), in
this paragraph we have demonstrated that the switch--on of a radio
pulsar (associated to a significant drop in mass transfer) could
determine ejection efficiencies close to $100\%$ as the matter is
ejected before it falls into the deep gravitational potential well of
the primary. 

For compact systems, the evolution outlined in Ic, IIc
implies that radio ejection is swiftly quenched by a resume of the
original mass--transfer rate, leading to the prediction that the amount
of mass accreted is substantial. On the other hand, if a radio
ejection starts in a wide system (IIw), IIIw implies that the
accretion is inhibited in the subsequent evolution. This suggests
that wide systems are the progenitors of the observed population of
moderately massive and moderately fast spinning MSPs.

\section{Where to search for ultra fast spinning NSs} 

The population synthesis calculations, described in \S 2, which include 
significant mass ejection and different assumptions for the evolution
of the magnetic field, showed that the process of recycling in LMXBs can 
produce a significant amount of ultra fast spinning objects. 
In line with this, a statistical analysis based on the current samples of 
detected MSPs, using different hypothesis for the period distributions of 
these sources, proved that there is always a non negligible probability 
for MSPs with $P<P_{\rm min}$ (Cordes \& Chernoff 1997). 
However, despite the large observational efforts over the recent years 
(D'Amico 2000, Edwards 2000, Crawford, Kaspi \& Bell 2000), no pulsar with 
$P<P_{\rm min}$ has been discovered so far. 
 
In \S 4 we noted that if the NS is spinning very fast, a drop in  
the mass transfer rate causes the switch--on of a radio pulsar and 
starts a radio ejection phase. Moreover, a new phase of accretion is only 
possible if the orbital period is short enough. Thus, our model predicts that 
the spin up of a NS to $P\lsim 1$ ms requires very close 
($P_{\rm orb}\lsim 1$ h) binary systems.

A consequence of the shortness of $P_{\rm orb}$ is that the strong
Doppler modulation of the pulsar period induced by the orbital motion
provide a natural observational bias against the detection of ultra fast 
spinning pulsars. 
All the algorithms for the detection of periodicities from a source in a 
close binary system are, unavoidably, a trade--off between computational capability and 
sensitivity: on each data set, they must perform a two dimensional search 
in the space of the dispersion measure DM (related to the unknown distance 
of the object), and acceleration (resulting from binary motion). 
Very short data sets reduce the Doppler modulations during the observation 
and reduce the CPU demands, at the price of limiting the sensitivity
(see {\it e.g.} Camilo \etal 2000). As a consequence, $P_{\rm orb}$ shorter 
than $\sim 90$ min have been poorly searched up to now, even in the most 
favorable case of searches on a definite target (such as those in
globular clusters), in which one of the parameters (DM) is known. 
The two MSPs with the shortest orbital periods, $96$ min and $102$ min, have 
been just detected in the globular clusters 47 Tuc (Camilo \etal 2000) and 
NGC 6544 (D'Amico \etal 2000), whereas in the galactic field 
the best case is that of PSR~J2051$-$0827 ($P_{\rm orb}=2.37$ h).
This observational bias would be worse in the presence of eclipses (favoured 
in very close binary pulsar systems, Nice 2000), in case of large duty 
cycles of the pulsed signal, in case of low radio luminosity, and strong 
interstellar scintillation, phenomena which have already been suggested to 
explain the elusiveness of SMSPs 
(Possenti 2000).  

\section*{Aknowledgments}
  
This work was supported by a grant by the Italian {\it Ministero dell' 
Universit\`a e della Ricerca Scientifica} (Cofin-99-02-02).


\begin{references}  
\small
Alpar, M.A., Cheng, A.F., Ruderman, M.A., \& Shaham, J. 1982, Nature,
300, 728


Bhattacharya, D. 1995, in X--ray Binaries, ed. W.H.G. Lewin, J. van
Paradijs \& E.P.J. van den Heuvel (Cambridge Univ. Press), 5


Burderi, L., D' Amico, N. 1997, ApJ, 490, 343

Burderi, L., Di Salvo, T., Robba, R., Del Sordo, S., Santangelo, A., 
Segreto, A. 1998, ApJ, 498, 831 

Burderi, L., King, A.R. \& Wynn, G.A. 1996, ApJ, 457, 348

Burderi, L., Possenti, A., Colpi, M., Di Salvo, T., \& D'Amico, N. 1999, 
ApJ, 519, 285

Camilo, F., Lorimer, D.R., Freier, P., Lyne, A.G., Manchester, R.N., 2000,
ApJ, 535, 975

Chen, K. \& Ruderman, M. 1993, ApJ, 402, 264

Cook, G.B., Shapiro S.L., \& Teukolsky, S.A. 1994, ApJL, 423, L117

Cordes, J. \& Chernoff, D.F. 1997, ApJ, 482, 971

Crawford, F., Kaspi, V.M. \& Bell, J.F.2000, in {\it Pulsar Astronomy - 
2000 and beyond}, ASP Conf.Series, 202, 31 

D'Amico, N. 2000, in {\it Pulsar Astronomy - 2000 and beyond}, 
ASP Conf.Series, 202, 27

D'Amico, N., Lyne, A.G., Manchester, R.N., Possenti, A., Camilo, F. 2000,
Nature, in press

D'Antona, F., Mazzitelli, I., Ritter, H. 1989, A\&A, 225, 391

Edwards, R.T., 2000, in {\it Pulsar Astronomy - 2000 and beyond}, 
ASP Conf.Series, 202, 33

Frank, J., King, A.R., Raine, D.J. 1992, {\it Accretion power in 
astrophysics}, second edition, Cambridge Astrophysics Series

Hayakawa, S., 1985, Phys. Rep., 121, 317

Illarionov, A., \& Sunyaev, R. 1975, A\&A, 39, 185

King, A.R., Kolb, U., \& Burderi, L. 1996, ApJ, 464, L127

Konar, S., \& Bhattacharya, D., 1999, MNRAS, 303, 588

Miri, M.J., \& Bhattacharya, D. 1994, MNRAS, 269, 455

Nice, D.J., Arzoumanian, Z., \& Thorsett, S.E. 2000, 
in {\it Pulsar Astronomy - 2000 and beyond}, ASP Conf.Series, 202, 67

Possenti, A., 2000, Ph.D. Thesis, Bologna, 159

Possenti, A., Colpi, M., D'Amico, N., \& Burderi, L. 1998, ApJL, 497, L97 

Possenti, A., Colpi, M., Geppert, U., Burderi, L., \& D'Amico, N. 1999, 
ApJS, 125, 463 

Ritter, H. 1988, A\&A, 202, 93

Ruderman, M., Shaham, J., \& Tavani, M. 1989, ApJ, 336, 507

Ruderman, M.A., Zhu, T.,\& Chen, K 1998, ApJ, 492, 267

Sang, Y., \& Chunmugam, G. 1987, ApJ, 323, L61

Shakura, N.I., , \& Sunyaev, R.A. 1973, A\&A, 24, 337

Srinivasan, G., {\it et al.} 1990, Curr.Sci. 59, 31 

Taam, R.E., King, A.R., \& Ritter, H. 2000, ApJ, submitted

Tanaka, Y., Shibazaki, N., 1996, ARA\&A, 34, 607

Tauris, T.M., \& Savonije, G.J. 1999, A\&A, 350, 928 

Tavani M. 1991, Nature, 351, 39

Thorsett, S.E., \& Chakrabarty, D. 1999, ApJ, 512, 288

Urpin, V.A., Geppert U., \& Konenkov, D. 1998, MNRAS, 295, 907

van der Klis, M. 1998, in {\it The Many Faces of Neutron Stars}, eds. Alpar,
Buccheri, van Paradijs, Dordrecht: Kluwer, 337

van Paradijs, J. 1996, ApJ, 464, L139

Webbink, R.F., Rappaport, S.A., \& Savonije, G.J. 1983, ApJ, 270, 678


\end{references}
\end{document}